# The Energy Landscape, Folding Pathways and the Kinetics of a[1] Knotted Protein

Michael C. Prentiss[1,2,*], David J. Wales[3], Peter G. Wolynes[1,2]

**1 Department of Chemistry, Center for Theoretical Biological Physics, University of California at San Diego, La Jolla, CA 92093, USA**
**2 International Institute for Complex Adaptive Matter, University of California, Davis, CA 95616 USA**
**3 University Chemical Laboratories, Lensfield Road, Cambridge CB2 1EW, United Kingdom**
∗ **E-mail: mcprentiss@gmail.com**

## Abstract

The folding pathway and rate coefficients of the folding of a knotted protein are calculated for a potential energy function with minimal energetic frustration. A kinetic transition network is constructed using the discrete path sampling approach, and the resulting potential energy surface is visualized by constructing disconnectivity graphs. Owing to topological constraints, the low-lying portion of the landscape consists of three distinct regions, corresponding to the native knotted state and to configurations where either the N- or C-terminus is not yet folded into the knot. The fastest folding pathways from denatured states exhibit early formation of the N-terminus portion of the knot and a rate-determining step where the C-terminus is incorporated. The low-lying minima with the N-terminus knotted and the C-terminus free therefore constitute an off-pathway intermediate for this model. The insertion of both the N- and C-termini into the knot occur late in the folding process, creating large energy barriers that are the rate limiting steps in the folding process. When compared to other protein folding proteins of a similar length, this system folds over six orders of magnitude more slowly.

## Author Summary

Proteins are chains, which must fold into a compact structure for the molecule to perform its biological function. There are a large number of ways the molecule can move into this final shape. Proteins have evolved sequences that perform this difficult task by having strong biases toward the final shape, while not getting stuck in different structures along the way. One way proteins can be trapped is by forming a knot in the chain. For the most part, proteins are remarkable in avoiding knotting. However, in order to function a few proteins form knots. We show how a model protein is able to knot itself, and estimate how fast this process occurs. Our goal is to treat a small and uncomplicated protein to estimate the fastest rate possible for the folding of a knotted protein. This rate is interesting when compared to the speed of folding of other proteins. We have visualized how the molecule changes shape to its functional position, and examined other paths the molecule may take.

## Introduction

The wide range of kinetics that characterizes protein folding has attracted interest from both experimentalists and theoreticians for decades [1]. Proteins fold on time scales that vary from microseconds to minutes [2], even though the corresponding energy landscape directs folding towards its native state. This wide range of rates can be explained by the diverse size and shapes of the free energy barriers between the unfolded and folded ensembles, which are largely determined by the pattern of contacts, often called the protein topology. Some of the slowest folding proteins, such as the green fluorescent protein, are both long and have complicated contact patterns [3]. In the past 15 years, a set of proteins has been discovered that fold slowly and have knotted topologies [4]. Topological constraints can lead to large energy barriers that are difficult to characterize, so we employ a reduced description of the protein and explore the landscape using geometry optimization techniques. Topological problems have been investigated previously using Gō models [5], and have shown that local unfolding may be required in some cases to organize the sequence of the folding of structural units. This type of potential energy landscape analysis with transition state theory has been used for describing kinetic phenomena in systems as diverse as molecular clusters, glasses, and proteins [6]. An advantage of this approach is that the topological constraint limits the utility of a simple energy based reaction coordinates (like $Q$, see Methods section), which ordinarily work well to describe folding.

We studied a tRNA methyltransferase (PDB code 1UAM), which contains a deep trefoil knot in the C-terminus domain [7]. The goal of this study was to estimate the fastest speed possible for folding a small knotted protein, and we therefore truncated the system to residues 78–135 to limit the number of atoms not included in the knot. Mathematically knots are defined in closed loops. In proteins links are used to connect the termini, and the structure is topologically classified by the determination of its Alexander polynomial [8,9]. Recently knotted proteins have been identified [10] and their kinetics explored. Knotted systems with a larger number of residues beyond the knot will exhibit slower kinetics, because of the need to break a larger number of contacts to fold properly. Protein models based on random contacts produce knotted proteins with greater frequency than is seen in protein structure databases [11, 12]. Knotted proteins are likely avoided during evolution, while some have remained and are an evolutionary curiosity. Knotting can also occur in other biopolymers such as DNA, and these systems exhibit significantly slower kinetics [13].

## Results

A connected path of minima and transition states between an unfolded structure and the native state was created with the discrete path sampling method (DPS) combined with the associative memory Hamiltonian [14, 15]. After obtaining an initial connection, this database of structures was expanded using schemes to systematically reduce the length and barriers associated with the largest contribution to the $k^{SS}$ rate coefficient [16–19]. Here, the SS superscript refers to an approximate formulation where the steady-state condition is applied to minima outside the product and reactant sets. This formulation provides a convenient framework for analysis because $k^{SS}$ can be written as a sum over discrete paths [16, 20]. Once this pathway appeared to have converged, the database was further refined by connecting undersampled minima with a large ratio of the free energy barrier to free energy difference from the global minimum [21, 22]. This choice is motivated by the idea of optimizing folding kinetics for topologically constrained system in a similar way to the minimization of energetic frustration obtained by comparing the folding temperature to the glass transition temperature [23, 24].

The resulting database contained 212054 minima and 206923 transition states, and the corresponding disconnectivity graph [25, 26] is shown in Figure 1, as rendered by VMD [27]. Here we remind the reader that every vertical line in the graph terminates at the energy of a local minimum, and that the minima are progressively connected together as the threshold energy, $E$, increases, according to the lowest barrier between them. The graph exhibits three distinct color coded features corresponding to potential energy basins, with properly knotted minima occupying the lowest-lying states in the center of the figure. Branches corresponding to minima with the knotted topology are colored blue, while those with the C and N termini still free are colored green and red, respectively. The kinetic coefficients for interconverting minima within these basins are relatively fast, so that local equilibrium is achieved on the time scale of the slow kinetics determined by the barriers between the different basins. This figure represents an unusual folding energy landscape, where large energy barriers occur despite the lack of favorable non-native interactions in the Hamiltonian. The two higher energy sets of structures correspond to local minima where either the C- or N-terminus has the native topology, but the other terminus is still unknotted, and we will refer to these as N-free and C-free geometries, respectively (Figures 2 and 3).

A useful descriptor of these ensembles is their structural overlap. The $Q$ value (a measure of structural similarity, see Methods section) between the N-free and native minima is 0.90, between the C-free and native minima is 0.87, and between N-free and C-free minima is 0.78. The small variation in $Q$ shows that only a few contacts are different, where a contact is restricted to be less than 9 Å in order to distinguish structures that have a high degree of similarity. Most of the contacts in each basin are identical, except for a few important differences near the termini. In the N-free minima these interactions are between residues 7-8 and 45-46 as shown in Figure 4, while in the C-free minima they are between 48-52 and 30-34 as shown in Figure 5, and define the interactions that prevent unphysical chain crossings. These non-native contacts are energetically neutral with respect to the interaction Hamiltonian $E_{int}$ (see Methods section) due to the native-only form of the Hamiltonian, but they affect the calculated pathways through an excluded volume repulsion due to the backbone interaction $E_{back}$.

A kinetic analysis of the DPS database using transition state theory requires choices for the value of $\epsilon$ and the mass associated with each site in the AMH potential. For simplicity a value of 12 atomic mass units (amu) was assigned to each site. To assign a value for $\epsilon$, we compared the normal mode frequencies for the AMH potential with typical values associated with heavy atom motion in all-atom representations of proteins. This comparison suggests that $\epsilon$ should be around one kcal/mol. The discrete path that makes the largest contribution to the phenomenological two-state rate coefficient, which we use to define the overall reaction mechanism, exhibits the same qualitative features over a wide range of values for $\epsilon$. The estimated rate coefficients themselves are more sensitive, as discussed below. If we take $\epsilon = 1$ kcal/mol and the values of length and mass in the AMH potential as 1 Å and 12 amu then the reduced value of $k_B T$ at room temperature is approximately 0.59. The pre-exponential factor for each minimum-to-minimum rate coefficient scales as $\sqrt{\epsilon}$, while the reduced value for room temperature decreases linearly, lowering the corresponding Boltzmann factors exponentially.

The choice of the reactant and product states can have a significant effect on the calculated rates. One way of selecting the states is to calculate an order parameter for all the local minima, and simply assign reactant and product states on this basis. However, an alternative method is possible using the known characteristics of the kinetic transition network and a self-consistency condition. Here we take the two endpoints that were used to calculate an initial path, and assign these minima to reactant and product sets. We then regroup the database using a recursive scheme [21], combining free energy minima that can interconvert without encountering a barrier higher than a chosen threshold value, $\Delta F_{barrier}$. This approach is attractive, because we require a separation of time scales for equilibration in the product and reactant regions, compared to the folding transition time, in order to recover a two-state description of the kinetics [20, 28, 29]. In this case, we expect to see a range of values for $\Delta F_{barrier}$ that give a similar value for the calculated rate coefficient.

Rate coefficients were calculated for three different choices of the reactants, namely a fully unfolded minimum and low-lying minima from the C-free and N-free regions of the landscape. In each case a low-lying minimum from the set of knotted configurations was chosen as the product, and rate coefficients were calculated for a range of $\epsilon$ and $\Delta F_{barrier}$ values. Following the recursive regrouping of states according to the value of $\Delta F_{barrier}$, mean first-passage times were calculated from each minimum in the reactant set using a graph transformation procedure [17, 30]. A phenomenological two-state rate coefficient is then obtained using appropriate occupation probabilities for the starting minimum [16, 17, 20, 30]. These values of $\epsilon$ tested (1.0, 0.9 and 0.7 kcal/mol) are close to the magnitude suggested by examining the normal mode frequencies. For $\epsilon = 1$ kcal/mol the rate coefficient varies from 0.04 to $0.4\,s^{-1}$ for $0 \leq \Delta F_{barrier} \leq 7$ kcal/mol using a fully unfolded or C-free minimum as the reactant. Therefore the folding time is between 25 seconds and 2.5 seconds.

A movie of the C-free minimum to the folded state is shown in Video S1. With a low-lying N-free minimum as the reactant, the calculated rate coefficient is $0.02\,s^{-1}$ for $0 \leq \Delta F_{barrier} \leq 4$ kcal/mol, jumping to about $50\,s^{-1}$ for $5 \leq \Delta F_{barrier} \leq 8$ kcal/mol. So the folding time becomes 50 seconds to 0.02 seconds. A movie of the N-free minimum to the folded state is shown in Video S2. The values obtained rise by about two orders of magnitude starting from N-free as reactant if we set $\epsilon = 0.7$ kcal/mol.

For each choice of reactant, the discrete path making the largest contribution to the $k^{SS}$ rate coefficient [16] was extracted, and snapshots of the intervening structures are superimposed upon the energy as a function of integrated path length in Figures 6, 7, and 8. Here the path length is defined from the Euclidean distance between successive configurations in the folding reaction. These pathways are based on the rate coefficients and associated free energy barriers calculated at 298 °K, and the entropy terms that derive from alternative discrete paths through the stationary point database are all included in the estimates of the overall rate coefficients. This kinetic analysis suggests that the local maxima in the energy profiles shown in Figures 6, 7, and 8 generally correspond to kinetic bottlenecks. Starting from an unfolded state, we see that the N-terminus first forms a loop that threads through the middle of the protein, and then opens (Figure 6). The structures involved in this process appear very similar to the corresponding event in the pathway starting from an N-free minimum in Figure 8, aside from the state of the C-terminus. The final folding events illustrated in Figure 6 are very similar to those shown in Figure 7, with a bend forming at the C-terminus, threading through the center of the structure, and straightening. The calculated rate coefficients are also very similar when the reactant is chosen as either the fully unfolded state or a C-free minimum, indicating that knotting of the C-terminus is the rate-determining step for this model system. The region of configuration space corresponding to low-lying N-free minima is then interpreted as a kinetic trap, which would probably result in a distinct relaxation time scale.

## Discussion

In this study a truncated sequence from a tRNA methyltransferase was considered with a Gō model containing only the favorable interactions that are present in the global minimum. In contrast to previous minimally frustrated models, which exhibit only a single potential energy funnel [31–33], the landscape for the knotted protein is divided into three distinct regions, corresponding to the correctly folded native state and to structures where either the C- or N-termini are not knotted. The potential energy barriers between the lowest minima in these regions are relatively large, with values of order 15 to 20 $\epsilon$ in units of the associated memory Hamiltonian [34, 35] parameter $\epsilon$, for which we estimate a value of around one kcal/mol. The calculated rates for folding are therefore rather slow, in agreement with previous simulations of knotted proteins [36, 37].

The folding reaction is hindered by the complex topology of this protein. Modeling these interactions and mechanisms in a realistic way requires new tools that prevent unphysical chain crossing events from occurring during the interpolation between structures that have an intervening knot. Details of the procedure are given in the Methods section, and an overview is provided here. Our initial aim was to avoid chain crossings by changing parameters of the potential. However, tightening the bond length constraints for covalent bonds does not solve the problem, because the interpolated images simply avoid the chain-crossing region. In the doubly-nudged [38] elastic band [39–41] (DNEB) method for identifying useful starting geometries for transition state refinement, a set of images are connected by harmonic springs, and the images can be equally spaced by increasing the corresponding force constant. However, this increase forces the chains into high energy structures that bracket an unphysical crossing. To avoid this situation it is necessary to construct a non-linear interpolation between the end points, and two strategies were implemented. To accelerate the energy evaluation, an elastic network potential was defined based on the two end points, with harmonic restraining potentials for atoms whose separation does not change. This geometrical analysis was also used to diagnose chain crossings for a linear interpolation between the end points, and to distinguish the chain that is moving from one that constitutes a barrier. When crossings were identified the potential was modified to shrink the end of the moving chain and add repulsive interactions to keep it away from the other chain. The DNEB images were then refined following standard procedures and the modified potential was morphed back into the AMH potential slowly enough for sites on one chain to move around the other chain, rather than through it. Overall, this procedure allows paths to be obtained that circumvent chain crossing, while retaining flexibility and providing a solution that is free of constraints.

Both the C- and N-termini must effectively cross over a chain belonging to the central region of the protein in order to achieve the knotted topology. In the present model it is the C-terminus crossing that appears to be the rate-determining step. The rate coefficients reported here are order of magnitude estimates, and correspond to a slow folding process, as expected from previous simulations [36, 37] and experiments [42, 43]. The precise energetics of this truncated model may differ from those of the full protein, but we expect the key steps in the folding and knotting pathways to be retained. Making a meaningful estimate of the scaling behavior with respect to chain length will be addressed in future work. For both the C- and N-termini the chain cross-over are achieved by formation of a loop, which then inserts through the center of the protein and straightens. While the chain is in the loop conformation the folding process could notionally be reversed by pulling the end of the chain, which is one definition of a slipknot topology, consistent with previous simulations [37]. The region of configuration space corresponding to N-free local minima, where the C-terminus crossing occurs first, is therefore likely to give rise to a separate relaxation time scale. The folding pathways exhibit some interesting mechanistic features, which might be transferable to related systems of knotted proteins and polymers. In particular, both the C- and N-termini crossings are achieved by formation of a loop that threads through the main body of the protein. The order of the knotting of the chain and the folding of the protein may change as the length of the system changes and as the energy function becomes more realistic. Adding non-native interactions would likely lower the free energy barrier of folding [44], but could also stabilize non-native structures and slow local refolding of the loops involved in chain crossings. Introducing non-additive cooperative contacts would increase the energy barriers and likely slow the kinetics [45]. Protein engineering studies of YibK suggest that knotting and formation of native structure are independent events that occur in sequence [46]. These experiments also suggest an early knotting event and slow development of native structure in the knotted region. Similar behavior is seen in DNA, where local unfolding speeds up diffusion of the knot along the polymer chain [47]. To provide meaningful comparisons with these observations will require simulations of longer systems, rather than the truncated sequence considered in the present work. When compared to other protein folding proteins of a similar length, which fold on the microsecond time scale [2], this system folds over six orders of magnitude more slowly.

## Methods

In order to describe an energy landscape with an exponential number of states, we reduce the atomistic detail of the system and discretize the energy landscape into minima and transition states. The associative memory Hamiltonian (AMH) protein model [34, 35], is a coarse-grained molecular mechanics potential inspired by the physics of protein folding. The energy functions consist of a polypeptide backbone term, $E_{\text{back}}$, with a molecular interaction term, $E_{\text{int}}$ [48–53]. The number of atoms per residue is limited to three ($C_{\alpha}$, $C_{\beta}$, and O), except for glycine. The interaction parameter $\epsilon$, which is the unit of energy, is defined by the native state energy excluding backbone contributions, $E_{\text{int}}$, via

$$\epsilon = \frac{|E_{\text{int}}|}{4N},\tag{1}$$

where $N$ is the number of residues. All temperatures are quoted in reduced units as $k_{\text{B}}T/\epsilon$. While $E_{\text{back}}$ creates self-avoiding peptide-like stereochemistry, $E_{\text{int}}$ introduces the majority of the attractive interactions that produce folding. Using the interactions described by $E_{\text{int}}$, we define a pairwise additive Gō model [54, 55], which is biased toward the native basin. Such models have been shown to reproduce many features of the mechanism and kinetics of protein folding [56, 57]. The interactions between residues were defined by,

$$E_{\text{int}} = -\frac{\epsilon}{a} \sum_{i < j-3} \gamma[x_{(|i-j|)}] \exp\left[-\frac{(r_{ij} - r_{ij}^{nat})^2}{2\sigma_{ij}^2}\right]$$ (2)

where the distances in the Gaussian term $r_{ij}^{nat}$ are determined by the native state. The interactions are defined in this minimal model for residues with greater than three residues sequence separation between the $C_\alpha - C_\alpha, C_\alpha - C_\beta, C_\beta - C_\alpha, C_\beta - C_\beta$ atom pairs. The weights, $\gamma$, corresponding to the depths of the Gaussian wells, are set to $(0.177, 0.048, 0.430)$ in order to approximately divide the interaction energy equally between the different distance classes, as suggested by previous theoretical models [58]. The width of the Gaussian, $\sigma_{ij}^2$, is determined by the sequence separation as $|i-j|^{0.15}$ Å. The scaling factor $a$ is used to satisfy Eq. (1). We measure the quality of the structures encountered with an order parameter, $Q$, which measures the sequence dependent structural similarity of two configurations. $Q$ is calculated from Eq. (3) as a summation of $C_\alpha$ pairwise differences between distances in a target and a reference structure (usually the native state), normalized by the number of contacts, where $N$ is sequence length:

$$Q = \frac{2}{(N-1)(N-2)} \sum_{i < j-1} \exp\left[-\frac{(r_{ij} - r_{ij}^{nat})^2}{\sigma_{ij}^2}\right].$$ (3)

The resulting order parameter, $Q$, ranges from zero, when there is no similarity between structures at a pair level, to unity, which indicates an exact overlap.

We made several changes to the original AMH backbone potential, $E_{\text{back}}$, in the present work. Eliminating some compromises necessary for rapid molecular dynamics simulations allows the AMH potential to be used with geometry optimization methods to produce tightly converged stationary points. This tight convergence is necessary for the construction of a kinetic transition network [20, 22, 59–61]. The terms shown in Eq. (4) are used to reproduce the peptide-like conformations in the original molecular dynamics energy function:

$$E_{\text{back}} = E_{\text{harm}} + E_{\text{Rama}} + E_{\text{chain}} + E_{\text{chi}} + E_{\text{ev}}.$$ (4)

For all calculations, we replaced the SHAKE method for bond constraints with a harmonic potential, $E_{\text{harm}}$, between the $C_\alpha$-$C_\alpha$, $C_\alpha$-$C_\beta$, and $C_\alpha$-O atoms. This replacement permits the location of local minima without requiring an internal coordinate transformation, and avoids discontinuous gradients [62]. The neighboring residues in sequence sterically limit the positions the backbone atoms can occupy, and this effect is reproduced with a Ramachandran potential, $E_{\text{Rama}}$. The planarity of the trans peptide bond is ensured by another harmonic potential, $E_{\text{chain}}$. The chirality of the $C_\alpha$ centers is maintained using the scalar triple product between neighboring $C_\beta$, C', and N atoms, $E_{\text{chi}}$. Excluded volume repulsion between the backbone atoms is achieved with via a smooth step (hyperbolic tangent) function, $E_{\text{ev}}$, in order to have a continuous potential, and differs from the previous hard sphere potential in the AMH.

For this Hamiltonian, we employed the discrete path sampling (DPS) approach to create databases of local minima and their intervening transition states, starting from two end points. To identify suitable endpoints, we used basin-hopping global optimization [63, 64] to search for the global minimum of the energy landscape, and to create an unfolded conformation. We have previously shown how basin-hopping can be successfully combined with associative memory Hamiltonians for identifying low energy states, and high quality structures [62]. The discrete path sampling approach is a coarse-grained analogue of the transition path sampling method [29, 65, 66], where geometry optimization tools are employed to refine a kinetic transition network. The network consists of local minima and transition states of the energy potential, where a transition state is defined as a stationary point with a single negative Hessian eigenvalue [67]. The connectivity is defined by approximate steepest-descent paths obtained by energy minimization following infinitesimal displacements parallel and anti-parallel to the eigenvector corresponding to the unique negative eigenvalue. A discrete path then refers to a series of minimum-to-minimum connections together with the intervening transition states. The original DPS formulation has been presented in detail elsewhere [16, 20], as have more recent developments [18, 21]. The aim is to enlarge a database of connected stationary points starting from those in the initial path, by adding all the minima and transition states found during successive connection-making attempts for pairs of minima selected from the current database.

The main challenge of DPS calculations is the characterisation of transition states. In contrast, energy minimization and identifying approximate steepest-descent pathways is straightforward; here we used the limited-memory Broyden–Fletcher–Goldfarb–Shanno (LBFGS) algorithm of Liu and Nocedal [68,69]. The transition state searches are connection attempts for a given pair of local minima. A doubly-nudged [38] elastic band [39–41] (DNEB) refinement of interpolated images was first run for each connection attempt, and the images corresponding to local energy maxima were then tightly converged using hybrid eigenvector-following [70]. The missing-connection algorithm [71] was employed to choose subsequent pairs of minima for further connection attempts [15].

To avoid unphysical chain crossed transition states, we made two changes in methodology to generate physical interpolations for finding potential transition state structures. We define two new potentials $V_{\mathrm{res}}$, which maintains chain connectivity, and $V_{\mathrm{rep}}$, which introduces atomic repulsion. The potential is modified in three stages during the DNEB refinement. During the first third of the DNEB steps $V_{\mathrm{res}}$ and $V_{\mathrm{rep}}$ are used with modified distances. In the second third the distances are relaxed to physically meaningful values, and in the final third we switch to the full AMH potential. We also define a simpler potential function, often referred to as an elastic network model [72] to represent the system during some of the DNEB refinement. The two end points for the DNEB calculation are analyzed to identify pairs of atoms within a cutoff distance ($10\,\mathrm{\AA}$) that are found at the same separation within a given tolerance in both structures. If $r_{ij}^S$ and $r_{ij}^F$ are the distances between atoms $i$ and $j$ in the starting and finishing geometries, then we introduced a harmonic restraining potential for this pair if $\left| r_{ij}^S - r_{ij}^F \right| / \bar{r}_{ij} < 0.1$, where $\bar{r}_{ij} = (r_{ij}^S + r_{ij}^F)/2$. For $N_{\mathrm{res}}$ such pairs the restraining potential was then

$$V_{\mathrm{res}} = A_{\mathrm{res}} \sum_{\alpha=1}^{N_{\mathrm{res}}} \left( r_\alpha - r_\alpha^0 \right)^2 , \qquad (5)$$

where $r_\alpha$ is the distance between the atoms involved in restraint $\alpha$, and $r_\alpha^0$ was initially set equal to $\bar{r}_\alpha = (r_\alpha^S + r_\alpha^F)/2$. The parameter $A_{\mathrm{res}}$ was set to $1000\,\epsilon$ in the present calculations, where the DNEB spring constant was set to $10\,\epsilon$. $V_{\mathrm{res}}$ has the appearance of an elastic network model [72], which reflects the conserved interatomic distances in the two endpoints. Analyzing the conserved distances is also useful for diagnosing when crossings occur, so that corresponding changes can be made to the potential, as described below. The initial images in the DNEB interpolation were simply placed at regular intervals for a linear interpolation between the specified endpoints, after putting these two structures into optimal alignment [73]. All pairs of atoms corresponding to different restraints with no common atoms were then examined for all pairs of DNEB images. The crossing check was applied for the largest untested image separation of every remaining image. Only pairs of restraints where the separation of the midpoints between the restrained atoms in both images were below a cutoff value of $10\,\mathrm{\AA}$ were considered. The midpoint separation in one of the two images was also required to change by at least $3\,\mathrm{\AA}$ from the value in the nearest endpoint structure. For restrained pairs satisfying these criteria a crossing is diagnosed when the dot product between the vectors joining the midpoints between the constraint pairs in the two images is negative. Outer and inner atom pairs are then defined according to how far the midpoints move between the two images: the midpoint that moves the furthest is assumed to belong to the outer chain, which needs to move around the inner chain.

To avoid unphysical crossings in the interpolation, we modify the potential $V_{\mathrm{res}}$ and add repulsive terms through $V_{\mathrm{rep}}$. If atomic contacts within the set of $N_{\mathrm{res}}$ pairs are found to cross, using the above geometrical condition, then repulsive terms are added according to the four distances between the two pairs of atoms. For $N_{\text{cross}}$ crossings of restrained distances, the repulsive contribution to the potential is

$$V_{\text{rep}} = B_{\text{rep}} \sum_{\gamma=1}^{N_{\text{cross}}} \sum_{\beta=1}^{4} \Theta(r_{\text{cross}}^{\text{cut}} - r_{\gamma,\beta})/r_{\gamma,\beta}, \tag{6}$$

where $\Theta$ is a step function, $r_{\text{cross}}^{\text{cut}}$ (10 Å) is a cut-off for the repulsive terms, $B_{\text{rep}}$ (100 $\epsilon$) defines the magnitude of the repulsion, and $r_{\gamma,\beta}$ is one of the four distances between pairs of atoms whose restrained contacts are found to cross. To enable chains to pass around one another when crossings are diagnosed, further changes were made to $V_{\text{res}}$. For the restrained contact in the outer and inner chains $r_{\alpha}^0$ was changed to $\max(r_{\alpha}^0/2, 0.01)$ and $\min(3r_{\alpha}^0/2, 2\bar{r}_{\alpha})$, respectively, for each crossing. Hence the outer chain shrinks while the inner chain expands. The first third of the DNEB iterations were run with the modified $V_{\text{res}}$ potential plus $V_{\text{rep}}$. For the middle third of the DNEB optimization the restraint distances $r_{\alpha}^0$ were switched back to the value $\bar{r}_{\alpha}$ according to the schedule $r_{\alpha}^0 \to (1-f)r_{\alpha}^0 + f\bar{r}_{\alpha}$, with $f = 10^{-4}$. The full AMH potential was then used for the last third of the DNEB iterations.

To describe the global kinetics of the transition network, we calculated the rate coefficients associated with each transition state using transition state theory [74] (TST) with vibrational densities of states obtained from harmonic normal mode analysis. The most important features of the mechanism of folding to the knotted state are relatively insensitive to the values assigned to minimum-to-minimum rate coefficients, while the total rate coefficients that we report are order of magnitude estimates.

## Acknowledgments

We thank Dr Joanne Carr and Dr Justin Bois for helpful comments.

# Figures

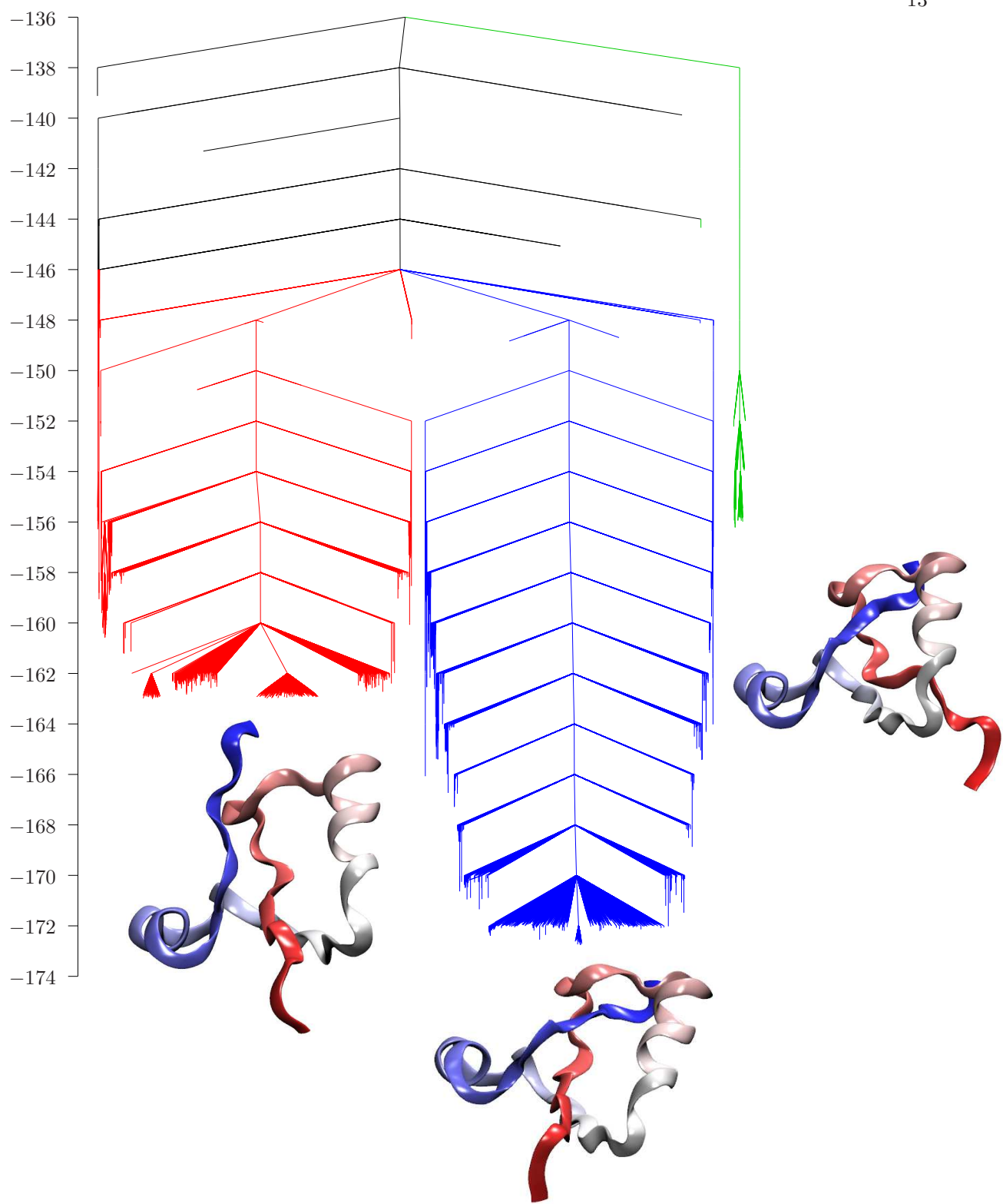

**Figure 1.** Disconnectivity graph for the truncated knotted protein with energy in $\epsilon$. This graph includes 10430 minima with three or more connections and 34519 transition states that link them. Branches corresponding to minima with the knotted topology are coloured blue, while those with the C and N termini still free are coloured green and red, respectively. The graph reveals a clear separation into three distinct features corresponding to the different topologies. The protein structures were rendered by VMD, shading the protein from blue to red between the N and C termini.

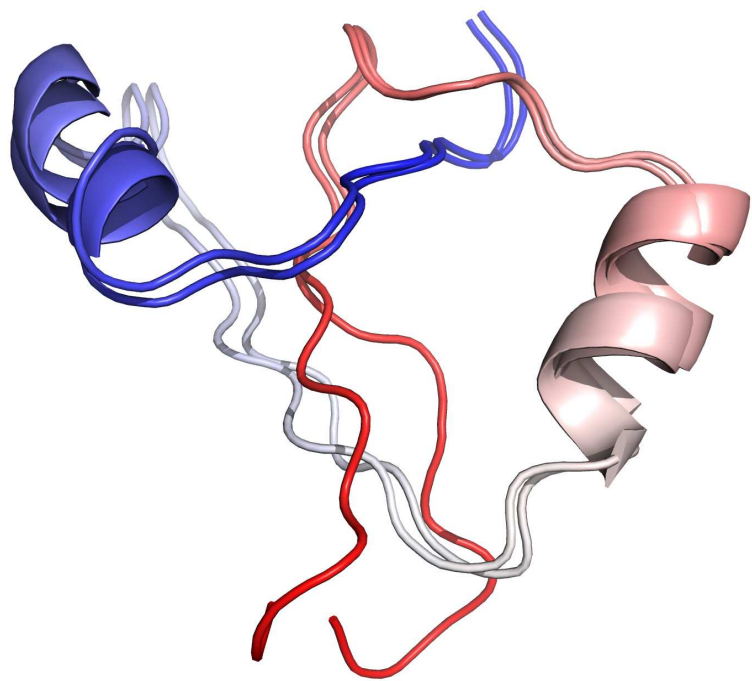

**Figure 2.** Structural overlap between the global minimum and a low-lying C-free structure. The native structure has the red C-free termini pointed towards the reader.

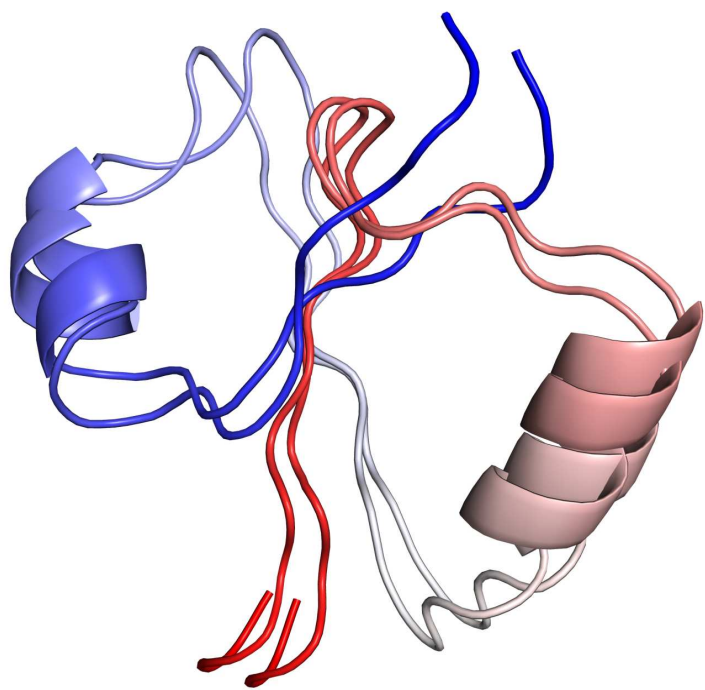

**Figure 3.** Structural overlap between the global minima and a low-lying N-free structure. The native structure has the blue N-free termini pointed towards into the paper.

Knotted Protein (PDB code 1UAM)

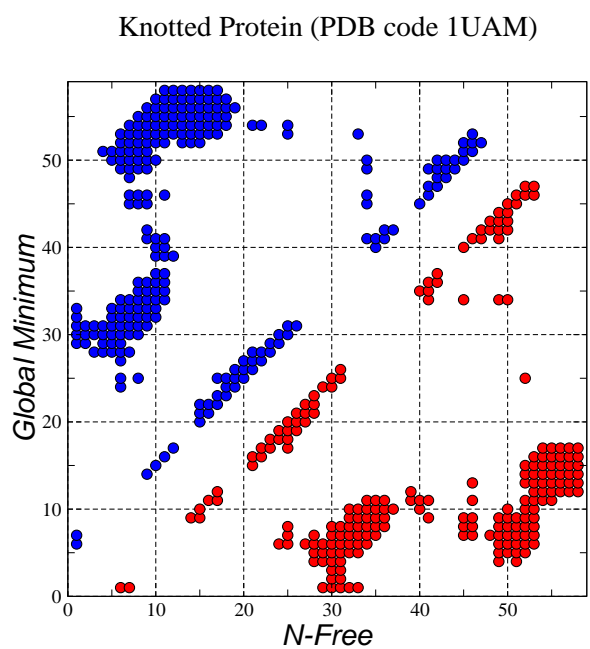

**Figure 4.** Contact map between the global minimum and the N-free structure.

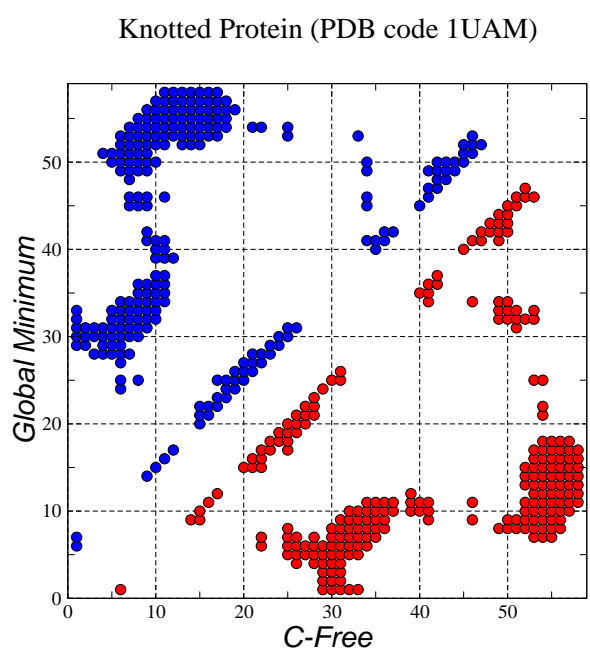

**Figure 5.** Contact map between the global minimum and the C-free structure.

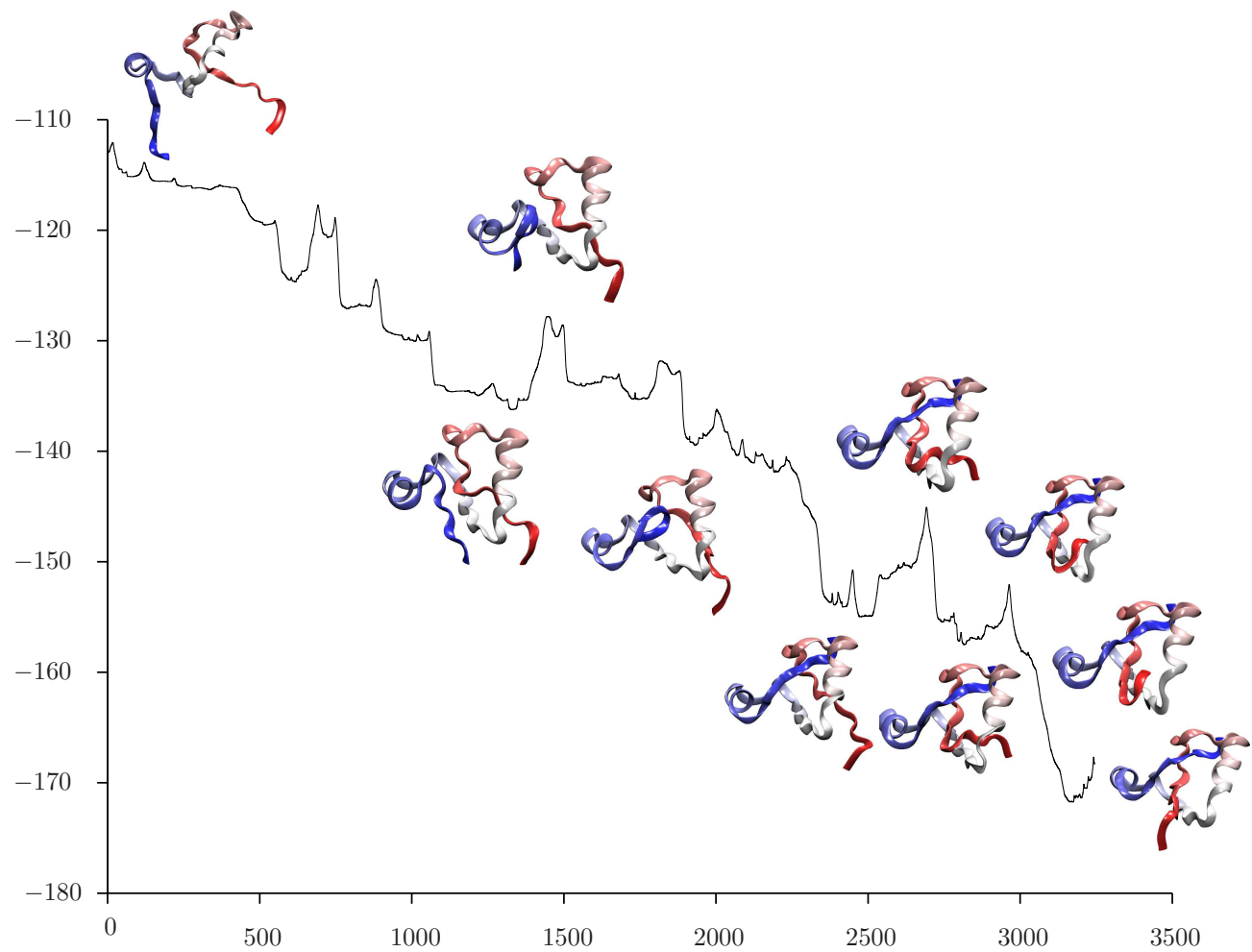

**Figure 6.** Energy/$\epsilon$ as a function of integrated path length/Å for the pathway that makes the largest contribution to the rate coefficient calculated for transitions from a denatured state to a low-lying knotted state. The integrated path length is the summation of the displacement of the atoms between the two end points.

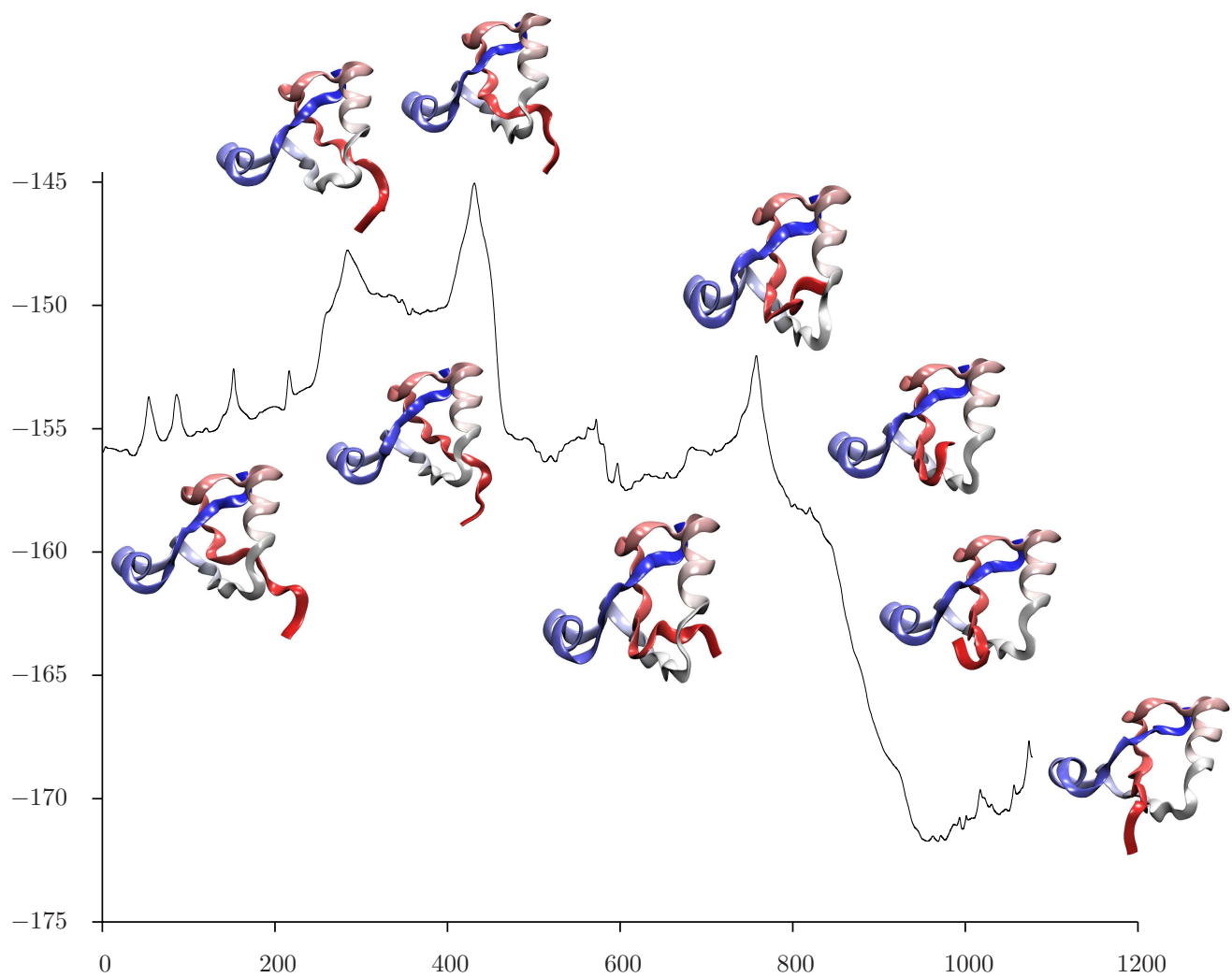

**Figure 7.** Energy/$\epsilon$ as a function of integrated path length/Å for the pathway that makes the largest contribution to the rate coefficient calculated for transitions from a low-lying minimum with a free C-terminus to a low-lying knotted state.

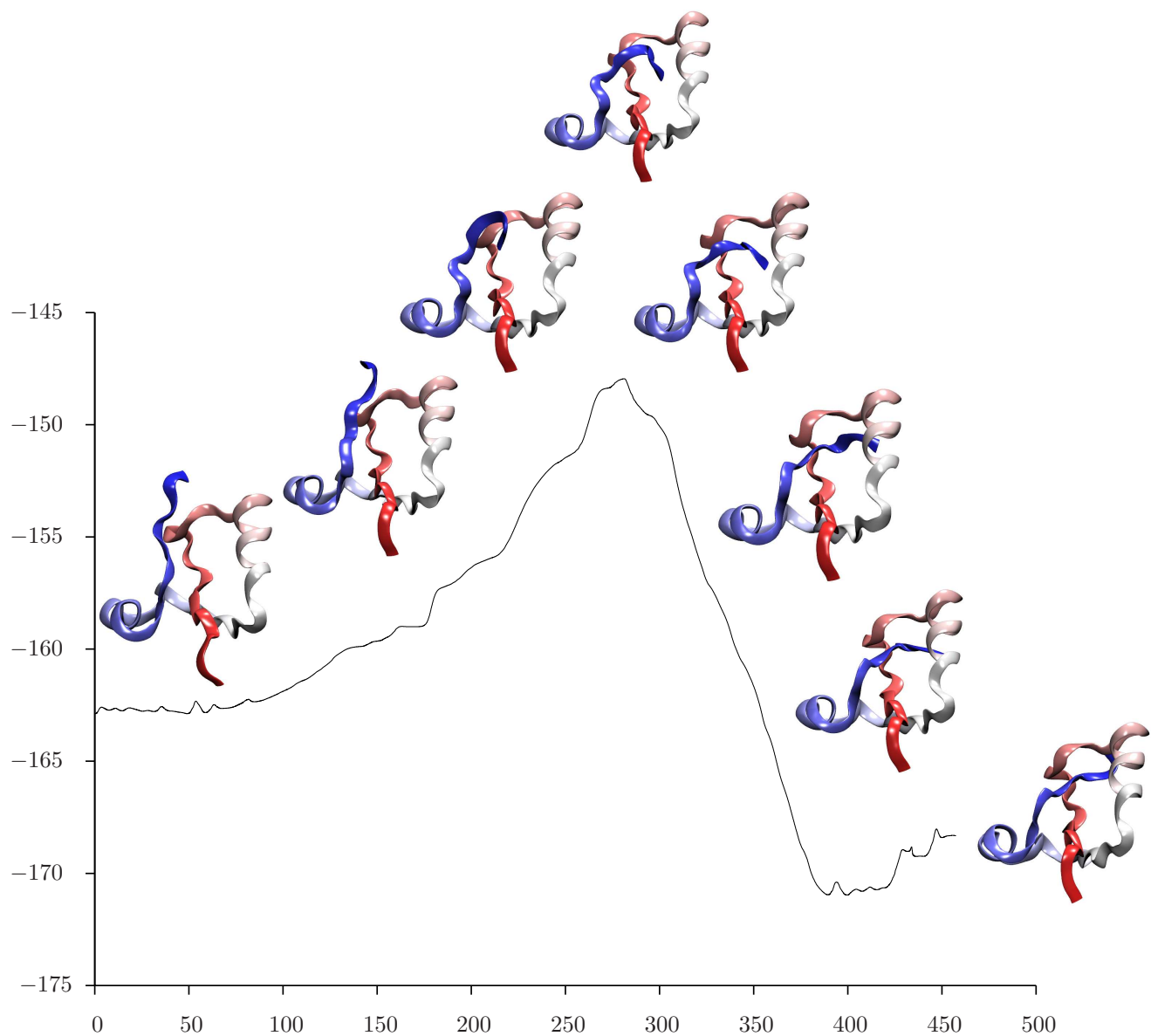

**Figure 8.** Energy/$\epsilon$ as a function of integrated path length/Å for the pathway that makes the largest contribution to the rate coefficient calculated for transitions from a low-lying minimum with a free N-terminus to a low-lying knotted state.